\documentclass[12pt]{iopart}
\usepackage{graphicx}
\usepackage{dcolumn}
\usepackage{bm}
\usepackage{iopams}
\usepackage{color}

\def\be{\begin{equation}}
\def\ee{\end{equation}}
\def\ba{\begin{array}}
\def\ea{\end{array}}
\def\bea{\begin{eqnarray}}
\def\eea{\end{eqnarray}}
\begin{document}
\title[\underline{}]
{ Half lives of proton emitters with KDE0v1 Skyrme interaction }
\author{K. Madhuri$^{1,2,*}$, D. N. Basu$^{3}$, T. R. Routray$^{1}$ and S. P. Pattnaik$^{1}$ }
\address{$^1$School of Physics, Sambalpur University, Jyotivihar-768019, India}
\address{$^2$Government Women's college, Sambalpur-768001, India}
\address{$^3$Variable Energy Cyclotron Centre, 1/AF Bidhan Nagar, Kolkata, 700064, India}

\ead{$^*$kmadhuriphy@gmail.com (corresponding author)}

\begin{abstract}
								The half-lives of proton radioactivity of proton emitters are investigated theoretically by using KDE0v1 Skyrme interaction.  The total barrier potential in the proton radioactive nuclei is calculated as the sum of the nuclear, Coulomb and centrifugal contributions. The Hartree-Fock nuclear density distributions is used in calculating the nuclear as well as the Coulomb interaction potentials. The quantum  mechanical tunneling probability is calculated within the WKB approximation. These calculations provide reasonable estimates for the observed proton radioactivity lifetimes.
\vspace{0.25cm}

\noindent
{\it Keywords}: Proton Radioactivity; KDE0v1 Skyrme effective interaction.

\end{abstract}

\maketitle

\section{Introduction}
		The nuclei lying beyond the proton drip line are proton rich, ({\it{i.e.}}, neutron deficient ) emit proton at their ground state. These proton unstable nuclei have positive Q-values.  This phenomenon of proton emission limits the possibilities of formation of more exotic nuclei on the proton rich side of the $\beta$ stability valley that is usually produced by fusion-evaporation nuclear reactions. The decaying proton is the unpaired proton not filling its orbit, hence it can be used as a tool to get spectroscopic informations. The decay rates are sensitive to the orbital angular momentum and the Q-values which in turns can be utilized to obtain the orbital angular momenta of the emitted protons. The proton decay is a nuclear reaction which can be written as 
					   $_{Z+1}^{A+1}X =_{Z}^{A}Y +_{1}^{1}H +Q_p$. 
        
Initially, the parent nucleus is in a quasistationary state and the proton may be considered as to tunnel through the potential barrier. In  the most cases, the combined Coulomb and centrifugal potentials give rise to barriers which are as large as $\sim$15 MeV. Consequently, the associated lifetimes, ranging from 1 $\mu$s to a few seconds, are very useful to give spectroscopic informations. So far the proton radio activity has been observed in different nuclei and their isotopes those are Sb, Tm, Lu, Ta, Ir, Au and Bi (spherical proton emitters) and of I, Cs, La, Pr, Eu, Tb and Ho (deformed proton emitters). The only three elements which are not in the list between Z=51 and Z=83, are Te(Z=52),Pm(Z=61)and Hg(Z=80). 
	Many theoretical attempts have been made to study the exotic nuclei, proton radioactivities and their half-lives \cite{BMP92,Ab97,Ab98,BA05,BCS05,De06,MG07,BCS08,YENI11,Fe11,TRR12}. In the present work, quantum mechanical tunneling probability is calculated within the WKB approximation using proton nucleus interaction potentials, as proton can be considered as a point charge having highest probability of being present in the parent nucleus in the energy domain of radioactivity. It has the lowest Coulomb potential among all charged particles and
mass being smallest it suffers the highest centrifugal barrier, allowing this process suitable to be dealt with WKB approximation. To obatin the potential barrier one requires the phenomenological parametrizations \cite{Gu99}, from fusion reaction studies \cite{BA05}, semiclassical considerations based on liquid drop model and the proximity potential \cite{Do09,Do10,Zh09,Zh10} and folding the nucleon-nucleon (N-N) effective interaction over the density distribution of the daughter nucleus \cite{BCS05,MG07,BCS08,YENI11}. The JLM \cite{MG07}, DDM3Y \cite{BCS08} and YENI \cite{YENI11} N-N effective interactions have been used in the folding model calculations.

      Eralier studies \cite{BV72,BV75,BV81} reveal that, Skyrme interaction gives very good results in finite nucleuse calculations. In ref.\cite{TRR12} various Skyrme forces have been used to calculate the proton radioactivity which have given very good results. In our present work we have used the only KDE0v1 Skyrme set \cite{bka2005} for the reason discussed in section 2. The formalism of the proton nucleus barrier potential  for Skyrme interaction in addition with penetration probability and half-life is presented in section 2. The results of half-lives of different proton radioactiv nuclei for KDE0v1 force is presented in section 3. In section 4, we have summarized briefly and concluded.

\section{Theoretical formalism}
		The half life of radioactive nuclei is given by 
		
\begin{equation}
 T_{1/2} = \frac{ln2}{\lambda}.
\label{seqn1}
\vspace{0.0cm}
\end{equation}
In the unified theory of proton radioactivity, the decay constant $\lambda$ is defined as
\begin{equation}
\lambda = \nu P S_p,
\label{seqn2}
\vspace{0.0cm}
\end{equation}

   where, $\nu$ as the assault frequency corresponding to the zero-point vibration energy $E_v$, $S_p$ is the spectroscopic factor and P is the penetration probability from tunnelling through the potential barrier. P can be calculated from the improved WKB formula \cite{Ke35} given by,
\begin{equation}
 P= \Big[ 1+\exp \Big\{ \frac{2}{\hbar} \int_{R_a}^{R_b} {[2\mu (U^p(r) - E_v - Q)]}^{1/2} dr \Big\} \Big]^{-1},
\label{seqn3}
\vspace{0.0cm}
\end{equation}
\noindent    
   where, $\mu = mM_d/M_A$  is the reduced mass of the proton-daughter nucleus system and m is the mass of proton (which is approximately the nucleonic mass),  $M_d$ and $M_A$ are the masses of the daughter nucleus and the parent nucleus respectively.  The turning points $R_a$ and $R_b$ are determined from the Center-of-mass energy $E_{CM}=Q$  by solving equation
\begin{equation}
 U^p(R_a) =E_v+Q = U^p(R_b),
\vspace{0.0cm}
\end{equation}
\noindent 
   The solution, in fact, provides 3 turning points. The fragments oscillate between 1st and 2nd turning points and tunnels via 2nd and 3rd turning points,$R_a$ and $R_b$. 
    We need the potential barrier $U^p(r)$ to calculate the penetration probability. Hence to know the half life of proton emitters we have to calculate the potential barrier.
    The total potential, $U^p(r)$, is given by
\begin{equation}
 U^p(r)= U_N^p(r)+U_{Coul}+\frac{{\hbar}^2}{2mr^2}[l(l+1)]
 \label{seqn4}
\vspace{0.0cm}
\end{equation}
\noindent    
		 where, $U_N^p(r)$ is the nuclear potential, $U_{Coul}$ is the Coulomb potential and $\frac{{\hbar}^2}{2mr^2}[l(l+1)]$ is the cetrifugal potential with $l{\hbar}$ being the orbital angular momentum of the emitted proton. The coulomb potential comprises of direct and exchange part i.e. $U_{Coul}=U_{Coul}^d+U_{Coul}^{ex}$, given as
\begin{equation}
U^d_{Coul}(r) = 4\pi e^2[\frac{1}{r}\int_0^r r'^2\rho_p(r') dr'+ \int_r^\infty r' \rho_p(r') dr'] 
 \label{seqn5}
\end{equation}
\begin{equation}
U^{ex}_{Coul}(r) = -e^2(\frac{3}{\pi})^\frac{1}{3} \rho_p^\frac{1}{3}(r) 
 \label{seqn6} 
\end{equation}
\noindent		
		 where, e is the electronic charge. To calculate the p-N nuclear potential $U_N^p(r)$  we need the energy density, $H(\rho_n,\rho_p)$, of a nucleus which is given by
\begin{equation}
 H(\rho_n,\rho_p) =  \frac{\hbar^2}{2m}(\tau_n + \tau_p) +  V,
  \label{seqn7}
\end{equation}
			 where, $\rho_{n}(\vec{r})$ and $\rho_{p}(\vec{r})$ are the density of the neutron and proton at postion $\vec{r}$ respectively. The first term on right hand side of the eq.(7) is the kinetic part of the energy density which comprises of the kinetic energy density of neutron $\tau_n(\vec{r})$ and that of proton $\tau_p(\vec{r})$ at postion $\vec{r}$ and the second term is the interaction part. We have used Skyrme interaction in our present work which is formulated as \cite{Ch97,Ch98,Ch981},
\begin{eqnarray}
 V =A(\rho_n,\rho_p) +  B_1\rho \tau  +  B_2(\rho_p \tau_n + \rho_n \tau_p) \nonumber \\
     \hspace{.5cm}+C [ (\nabla\rho_n)^2 + (\nabla\rho_p)^2 ] + D (\nabla\rho_n)(\nabla\rho_p), 
\label{seqn8}
\end{eqnarray}
\noindent     
		with $\rho(\vec{r})=\rho_n(\vec{r})+\rho_p(\vec{r})$ and $\tau(\vec{r})=\tau_n(\vec{r})+\tau_p(\vec{r})$ being the total nucleonic density and total kinetic energy density, respectively, at position $\vec{r}$ and 

\begin{eqnarray}
 A(\rho_n,\rho_p) = \frac{t_0}{4} [(1-x_0)(\rho_n^2+\rho_p^2)+(4+2x_0)\rho_n \rho_p]  \nonumber \\ 
 \hspace{2cm}+\frac{t_3}{24}\rho^\gamma [(1-x_3) (\rho_n^2+\rho_p^2) + (4+2x_3)\rho_n\rho_p],   
 \label{seqn9} \\
\hspace{1cm} B_1 = \frac{1}{8}[t_1(1-x_1)+3t_2(1+x_2)],    \label{seqn10}  \\
\hspace{1cm} B_2 = \frac{1}{8}[t_1(1+2x_1)-t_2(1+2x_2)],   \label{seqn11}  \\
 \hspace{1cm}C = \frac{3}{32}[ t_1(1-x_1)-t_2(1+x_2)],     \label{seqn12}  \\
 \hspace{1cm}D = \frac{1}{16}[3 t_1(2+x_1)-t_2(2+x_2)].  \label{seqn13}
\end{eqnarray}
\noindent  
The real part of the p-N potential is the functional derivative of energy density $H(\rho_n,\rho_p)$ with respect to proton, i.e., $\frac{\partial H}{\partial [f]_p}$. Further the non-local Skyrme Hartree-Fock potential, {\it i.e.} specified by a nucleon effective mass, can be taken care by an equivalent energy dependent local potential \cite{Do72}. Considering this energy dependence of the real part of the nucleon-nucleus potential, the p-N potential $U_N^p(r)$ can now be given by
\begin{eqnarray}
 U_N^p(r) = \Big[ 1-\Big(\frac{m^*(r)}{m}\Big)_p \Big] (E_{CM}-U_{Coul})   \nonumber  \\
 \hspace{1.5cm}+ \Big(\frac{m^*(r)}{m}\Big)_p \Big[ \frac{\partial A(\rho_n,\rho_p)}{\partial \rho_p} + B_1 \tau + B_2 \tau_n - 2 C \nabla^2 \rho_p  \nonumber  \\
\hspace{1.5cm}- D \nabla^2 \rho_n + \frac{1}{2} \Big(\frac{d^2}{dr^2}\frac{\hbar^2}{2m^*_p(r)}\Big) - \frac{m^*_p(r)}{2\hbar^2}\Big(\frac{d}{dr}\frac{\hbar^2}{2m^*_p(r)}\Big)^2 \Big] 
 \label{seqn14}
\end{eqnarray}
\noindent
  where, the functional $A(\rho_n,\rho_p)$ and the functions $B_1$, $B_2$, $C$ and $D$ are given in eqs.(9)-(13) and $\left(\frac{m^*(r)}{m}\right)_p$ is the proton effective mass which is given by

\begin{equation}
 \left(\frac{m^*(r)}{m}\right)_p = \Big[ 1+\frac{2m}{\hbar^2} (B_1 \rho + B_2 \rho_n) \Big]^{-1}.
 \label{seqn15}
\vspace{0.0cm}
\end{equation}
\noindent  
In deriving eq.(4) we have used the fact that the wave number, $k$, of the proton at a distance $\vec{r}$ from the center of the nucleus taken as the origin is

\begin{equation}
 k(\vec{r}) = \Big[\frac{2m}{\hbar^2} \Big\{ E_{CM}-U_N^p(r)-U_{Coul} \Big\} \Big]^{1/2}.
 \label{seqn16}
\vspace{0.0cm}
\end{equation}
\noindent                      
 Now the proton-Nucleus potential $U_N^p(r)$ for the Skyrme interaction as derieved from eq.(10) using eqs.(9)-(13) is given by 
 
 \begin{eqnarray}
 &&U_N^p(r) = \Big[1-\Big(\frac{m^*(r)}{m}\Big)_p\Big] (E_{CM}-U_{Coul}) \nonumber  \\
 &&\hspace{1.5cm}+ \Big(\frac{m^*(r)}{m}\Big)_p \Big[ \frac{t_0}{2} [ (1-x_0)\rho_p+(2+x_0)\rho_n] \nonumber \\ 
 &&\hspace{1.5cm}+ \frac{t_3}{12}\Big( \gamma [(1-x_3) \frac{\rho_p^2+\rho_n^2}{2} + (2+x_3)\rho_p\rho_n] \nonumber \\
 &&\hspace{1.5cm}+\rho [(1-x_3)\rho_p+(2+x_3)\rho_n] \Big) \rho^{\gamma-1} \nonumber \\
 &&\hspace{1.5cm}+\frac{1}{8}[t_1(2+x_1)+t_2(2+x_2)]\tau_n \nonumber \\
 &&\hspace{1.5cm}+ \frac{1}{8}[t_1(1-x_1)+3t_2(1+x_2)]\tau_p \nonumber \\
 &&\hspace{1.5cm}-\frac{3}{16}[t_1(1-x_1)-t_2(1+x_2)](\nabla^2\rho_p) \nonumber \\
 &&\hspace{1.5cm}-\frac{1}{16}[3t_1(2+x_1)-t_2(2+x_2)](\nabla^2\rho_n) \nonumber \\
 &&\hspace{1.5cm}+\frac{1}{16}\Big\{[t_1(2+x_1)+t_2(2+x_2)]\frac{d^2\rho_n}{dr^2} \nonumber \\
 &&\hspace{1.5cm}+ [t_1(1-x_1)+3t_2(1+x_2)]\frac{d^2\rho_p}{dr^2}\Big\} \nonumber \\ 
 &&\hspace{1.5cm}-\frac{m^*_p(r)}{128\hbar^2}\Big\{[t_1(2+x_1)+t_2(2+x_2)]\frac{d\rho_n}{dr} \nonumber \\
 &&\hspace{1.5cm}+ [t_1(1-x_1)+3t_2(1+x_2)]\frac{d\rho_p}{dr}\Big\}^2 \Big].
 \label{seqn17}
\end{eqnarray}
\noindent
  The calculation of proton-Nucleus potential in the above equation requires the knowledge of nucleonic density distribution in the nucleus, their gradients and the kinetic energy densities. In our presnt work we have taken the Hartree-Fock (HF) density distribution for nucleonic density distribution $\rho_{n(p)}(\vec{r})$. The kinetic energy density $\tau_{n(p)}$ is taken to be the Thomas-Fermi one along with the second-order correction
 \begin{equation}
 \tau_{n(p)} = \frac{3}{5} k^2_{n(p)}\rho_{n(p)} + \frac{1}{36} \frac{(\nabla\rho_{n(p)})^2}{\rho_{n(p)}}+ \frac{1}{3}\nabla^2\rho_{n(p)},
\label{seqn18}
\vspace{0.0cm}
\end{equation}
\noindent       
   where, $k_{n(p)} = [ 3\pi^2 \rho_{n(p)} ]^{1/3}$,  is the neutron (proton) Fermi momentum at density $\rho_{n(p)}(\vec{r})$.
  The reason of using only the KDE0v1 skyrme interaction in the present text is that in ref. \cite{dutra12}, 240 skyrme interaction parameter sets has been examined by Dutra et al. by comparing their output in eleven microscopic constraints derieved mainly from experimental data and emperical properties of symmetric nuclear matter at and close to satuaration. They have found that only 5 of the 240 forces analyzed satisfy all the constraints imposed. These 5 types of Skyrme models are collectively named as CSkP* set, which includes KDE0v1, LNS, NRAPR, SKRA, and SQMC700. Further, constraints like maximum mass and the corresponding central density of high-mass neutron stars restrict the Skyrme model to describe the NS structure because it requires extrapolation to densities above the valid range. It has been proposed that the Tolman VII EOS-independent analytic solution of Einstein's equations marks an upper limit on the ultimate density of observable cold matter. If this argument is correct, it follows that mass measurement sets an upper limit on this maximum density of 10 times the saturation density \cite{demorest2010}. None of the CSkP* models produces a maximum mass neutron star with central density in line with observation except KDE0v1 and NRAPR parameter set. But NRAPR parameter set does not satisfy all the other constraints of nuclear matter. Using the KDE0v1 parameter set the maximum mass has been obtained corresponding to the central density  7.7 times the saturation density \cite {km17} and the KDE0v1 set tackles all the constraints as described in ref \cite{dutra12} and can explain NS structure. Moreover, only the KDE0v1 force can also fit a significant amount of data from finite nuclei quite well \cite{stone2013}.  Therefore we want to investigate the result of Proton radioactivity using KDE0v1 skyrme interaction.
\begin{table}
\centering
\caption{Values of the nine parameters of Asymmetric Nuclear Matter for the Skyrme interaction corresponding to KDE0v1 \cite {bka2005}. The nuclear saturation density $\rho_0$, the saturation energy per particle $e(\rho_0)$, the incompressibility of isospin symmetric nuclear matter $K(\rho_0)$, the effective mass $m^*$,  the nuclear symmetry energy at saturation density $E_s(\rho_0)$ and its slope $L$ for the Skyrme interaction corresponding to KDE0v1 are also provided.}
\renewcommand{\tabcolsep}{0.05cm}
\renewcommand{\arraystretch}{1.2}
\begin{tabular}{|c|c|c|c|c|c|c|c|c|c|c|c|}\hline
$\gamma$ & $t_0$ & $x_0$ & $t_1$ & $x_1$ & $t_2$ & $x_2$ & $t_3$ & $x_3$\\\hline
0.17&-2553.1 & 0.65&411.7&-0.35 &-419.9&-0.93&14603.6&0.95 \\\hline
\multicolumn{9}{|c|}{Nuclear matter properties at saturation density} \\
\hline
\multicolumn{1}{|c|}{$\gamma$}&\multicolumn{1}{|c|}{$\rho_0$ ($\mathrm{fm}^{-3}$)} & \multicolumn{2}{c|}{$e (\rho_0) $ (MeV)}
& \multicolumn{1}{c|}{$K (\rho_0)$ (MeV)} & \multicolumn{1}{c|}{$\frac{m^*}{m}(\rho_0,k_{f_0})$}
& \multicolumn{1}{c|}{$E_s (\rho_0)$ (MeV)} & \multicolumn{2}{c|}{$L (\rho_0)$ (MeV)} \\
\hline
\multicolumn{1}{|c|}{0.17}& \multicolumn{1}{|c|}{0.165} & \multicolumn{2}{c|}{-16.0} & \multicolumn{1}{c|}{227.54}
& \multicolumn{1}{c|}{0.74} & \multicolumn{1}{c|}{34.58} & \multicolumn{2}{c|}{54.69} \\\hline
\end{tabular}
\end{table}
   
\section{Results and discussion}
The p-N potential is calculated using KDE0v1 Skyrme interaction. The Values of the nine parameters of Asymmetric Nuclear Matter for the Skyrme interaction corresponding to KDE0v1 \cite {bka2005} and Nuclear matter properties at saturation density are given in Table.1. Hartree-Fock density distribution is used in calculation of the nuclear as well as the Coulomb interaction potentials. Fig.-1 represents the Hartree-Fock proton and neutron density distributions as functions of distance $r$ for $^{158}$W nucleus (daughter of $^{159}$Re proton emitter for KDE0v1 skyrme interaction). The sum of the centrifugal potential and the coulomb potential with the $U_N^p$ gives the total barrier potential $U^p(r)$, experienced by the emitting proton. The centrifugal potential has been calculated by using the orbital angular momentum (l) values given in ref. \cite{So02}. The pentration probability is calculated by using the eq.(3) for KDE0v1 Skyrme interaction. The experimental Q-values together with their uncertainties has been cosidered in the calculation. Then the half-lives of different proton radioactive nuclei is calculated using eq.(1), which is listed in Table.2. Table.2 includes half-lives of 31 cases of proton emitters. By using the current model, the measured half-lives of proton-rich nuclei have been compared with the experimental data along with the results of the folding model potential using DDM3Y effective interaction \cite{BCS08}. The present results are in good agreement with the existing data. There are some discrepancy between the measured half-life values and experimental data for some cases may be because of the uncertainty measurement in the Q-values. We found that the KDE0v1 skyrme interaction gives very good justice to proton radio activity for proton-rich nuclei.     
%
\newpage     
\begin{figure}[ht]
\vspace{0.0cm}
\begin{center}
\includegraphics[width=.8\columnwidth]{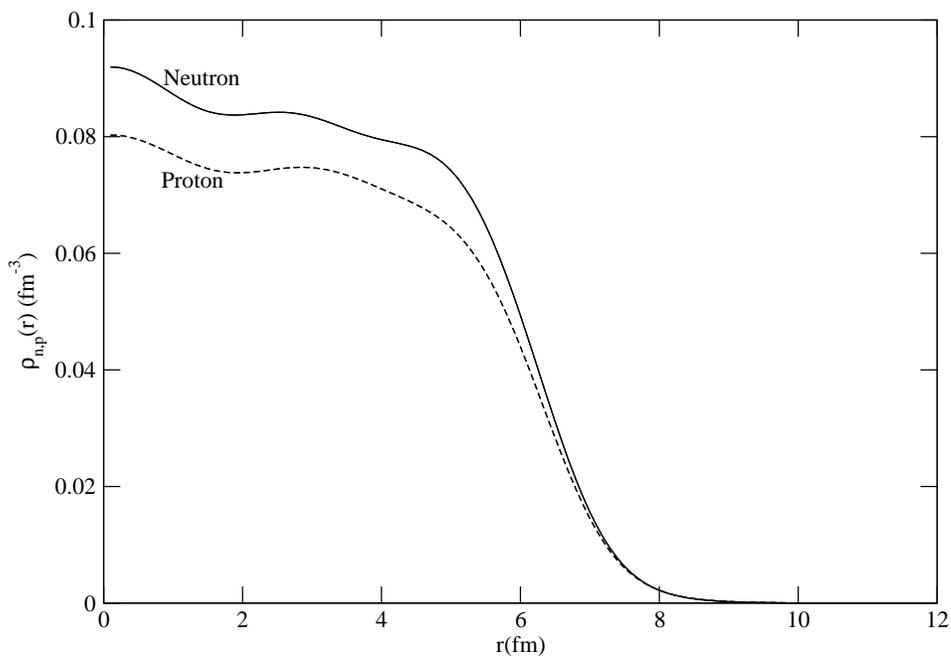}
\caption{KDE0v1 Hartree-Fock proton and neutron density distributions as functions of distance $r$ for $^{158}$W nucleus (daughter of $^{159}$Re proton emitter)}
\label{Figure.1}
\end{center}
\end{figure}
    
\newpage  
\begin{table*}[ht]
\vspace{0.0cm}
\caption{ The results of the present calculations using the Skyrme KDE0v1 p-N potentials are compared with the experimental values along with the results of DDM3Y \cite{BCS08}. Except $^{135}Tb$ \cite{Wo04} and $^{159}Re$ \cite{Jo06}, all the experimental $Q$ values, half lives and $l$ values are from Ref.\cite{So02}. Experimental errors in $Q$ values and corresponding errors in calculated half lives are inside parentheses. Asterisk symbol in the parent nucleus denotes isomeric state. The superscript $HF$ denotes results calculated with HF densities, its derivatives and kinetic energy densities.}
\vspace{0.0cm}
\center
\begin{tabular}{cccccccccccccc}
\hline
\hline
Parent & $l$ & $Q^{ex}$ &Measured&KDE0v1$^{HF}$&DDM3Y    \\ 
$^A Z$& $\hbar$ & MeV &$log_{10}T(s)$&$log_{10}T(s)$&$log_{10}T(s)$ \\ 
\hline
$^{105}Sb$&2&0.491(15)&2.049$^{+0.058}_{-0.067}$&1.98(46)&1.90(45) \\ 
$^{109}I$&2&0.829(3)&-3.987$^{+0.020}_{-0.022}$&-4.20(5)&-4.31(5) \\ 
$^{112}Cs$&2&0.824(7)&-3.301$^{+0.079}_{-0.097}$&-3.13(11)&-3.21(11) \\ 
$^{113}Cs$&2&0.978(3)&-4.777$^{+0.018}_{-0.019}$&-5.52(4)&-5.61(4) \\ 
$^{145}Tm$&5&1.753(10)&-5.409$^{+0.109}_{-0.146}$&-5.33(6)&-5.28(7) \\ 
$^{147}Tm$&5&1.071(3)&0.591$^{+0.125}_{-0.175}$&0.76(4)&0.83(4) \\ 
$^{147}Tm^*$&2&1.139(5)&-3.444$^{+0.046}_{-0.051}$&-3.45(6)&-3.46(6) \\ 
$^{150}Lu$&5&1.283(4)&-1.180$^{+0.055}_{-0.064}$&-0.79(4)&-0.74(4) \\ 
$^{150}Lu^*$&2&1.317(15)&-4.523$^{+0.620}_{-0.301}$&-4.44(15)&-4.46(15) \\ 
$^{151}Lu$&5&1.255(3)&-0.896$^{+0.011}_{-0.012}$&-0.88(3)&-0.82(4) \\ 
$^{151}Lu^*$&2&1.332(10)&-4.796$^{+0.026}_{-0.027}$&-4.94(10)&-4.96(10) \\
$^{155}Ta$&5&1.791(10)&-4.921$^{+0.125}_{-0.125}$&-4.86(7)&-4.80(7) \\ 
$^{156}Ta$&2&1.028(5)&-0.620$^{+0.082}_{-0.101}$&-0.45(8)&-0.47(8) \\ 
$^{156}Ta^*$&5&1.130(8)&0.949$^{+0.100}_{-0.129}$&1.44(10)&1.50(10) \\ 
$^{157}Ta$&0&0.947(7)&-0.523$^{+0.135}_{-0.198}$&-0.48(12)&-0.51(12) \\
$^{160}Re$&2&1.284(6)&-3.046$^{+0.075}_{-0.056}$&-3.05(7)&-3.08(7) \\ 
$^{161}Re$&0&1.214(6)&-3.432$^{+0.045}_{-0.049}$&-3.49(7)&-3.53(7) \\ 
$^{161}Re^*$&5&1.338(7)&-0.488$^{+0.056}_{-0.065}$&-0.80(8)&-0.75(8) \\ 
$^{164}Ir$&5&1.844(9)&-3.959$^{+0.190}_{-0.139}$&-4.12(6)&-4.08(6) \\ 
$^{165}Ir^*$&5&1.733(7)&-3.469$^{+0.082}_{-0.100}$&-3.72(5)&-3.67(5) \\
$^{166}Ir$&2&1.168(8)&-0.824$^{+0.166}_{-0.273}$&-1.17(11)&-1.19(10) \\ 
$^{166}Ir^*$&5&1.340(8)&-0.076$^{+0.125}_{-0.176}$&0.01(9)&0.06(9) \\ 
$^{167}Ir$&0&1.086(6)&-0.959$^{+0.024}_{-0.025}$&-1.31(9)&-1.35(8) \\ 
$^{167}Ir^*$&5&1.261(7)&0.875$^{+0.098}_{-0.127}$&0.48(8)&0.54(8) \\ 
$^{171}Au$&0&1.469(17)&-4.770$^{+0.185}_{-0.151}$&-5.05(16)&-5.10(16) \\ 
$^{171}Au^*$&5&1.718(6)&-2.654$^{+0.054}_{-0.060}$&-3.25(5)&-3.19(5) \\ 
$^{177}Tl$&0&1.180(20)&-1.174$^{+0.191}_{-0.349}$&-1.33(26)&-1.44(26) \\ 
$^{177}Tl^*$&5&1.986(10)&-3.347$^{+0.095}_{-0.122}$&-4.63(7)&-4.64(6) \\ 
$^{185}Bi$&0&1.624(16)&-4.229$^{+0.068}_{-0.081}$&-5.52(13)&-5.53(14) \\ 
$^{135}Tb$&3&1.188(7)&-3.027$^{+0.131}_{-0.116}$&-3.98(8)&-4.16(8) \\
$^{159}Re$&5&1.816(20)&-4.678$^{+0.076}_{-0.092}$&-4.64(13)&-4.59(13) \\ \hline
\hline
\end{tabular} 
\vspace{0.0cm}
\end{table*}
\eject  
\section{Summary and Conclusion}
  The purpose of this work is to provide a theoretical description of the proton radioactivity using KDE0v1 Skyrme interaction that satisfies all the constraints ranging from finite nuclei to neutron stars. Present calculations (vide Table.2) demonstrate that this framework is adequate to deliver an overall account of proton radioactivity half lives. The results of the present calculations are in good agreement over a wide range of experimental data. Moreover, only the KDE0v1 Skyrme interaction can also provide quite good description of a significant amount of data from finite nuclei as well \cite{stone2013}. This interaction addresses all the constraints described by Dutra et al. \cite{dutra12} and produces good results for mass-radius relation in Vela pulsar \cite{km17}. Present calculations reveal that the KDE0v1 Skyrme interaction also provides reasonably good description of proton radioactivity.
\newpage
\section*{References}



\begin{thebibliography}{99}
\bibitem{BMP92} B. Buck, A. C. Merchant and S. M. Perez, Phys. Rev. {\bf C 45}, 1688 (1992).
\bibitem{Ab97} S. Aberg, P. B. Semmes and W. Nazarewicz, Phys. Rev. {\bf C 56}, 1762 (1997). 
\bibitem{Ab98} S. Aberg, P. B. Semmes and W. Nazarewicz,Phys. Rev. {\bf C 58}, 3011 (1998).
\bibitem{BA05} M. Balasubramaniam and N. Arunachalam, Phys. Rev. {\bf C 71}, 014603 (2005). 
\bibitem{BCS05} D. N. Basu, P. Roy Chowdhury and C. Samanta, Phys. Rev. {\bf C 72}, 051601(R) (2005).
\bibitem{De06} D. S. Delion, R. J. Liotta and R. Wyss, Phys. Rep. {\bf 424}, 113 (2006) and references therein.
\bibitem{MG07} M. Bhattacharya and G. Gangopadhyay, Phys. Lett.  {\bf B 651}, 263 (2007).
\bibitem{BCS08} D. N. Basu, P. Roy Chowdhury and C. Samanta, Nucl. Phys. {\bf A 811}, 140 (2008).
\bibitem{YENI11} T. R. Routray, S. K. Tripathy, B. B. Dash, B. Behera and D. N. Basu, Eur. Phys. J. {\bf A 47}, 92 (2011).
\bibitem{Fe11} L. S.Ferreira E. Maglione and P. Ring, Phys. Lett. {\bf B 701}, 508 (2011).
\bibitem{TRR12} T. R. Routray, A. Mishra, S. K. Tripathy, B. Behera and D. N. Basu, Eur. Phys. J. {\bf A 48}, 77 (2012).
\bibitem{Gu99} F. Guzm\'an, M. Goncalves, O. A. P. Tavares, S. B. Duarte, F. Garc\'ia and O. Rodr\'iguez,
Phys. Rev. {\bf C 59}, R2339 (1999).
\bibitem{Do09} J. M. Dong, H. F. Zhang and G. Royer, Phys. Rev. {\bf C 79}, 054330 (2009).
\bibitem{Do10} J.-M. Dong, H.-F. Zhang, W. Zuo and J.-Q. Li, Chin. Phys. {\bf C 34}, 182 (2010).
\bibitem{Zh09} H. F. Zhang, J. M. Dong, Y. J. Wang, X. N. Su, Y. J. Wang, L. Z. Cai, T. B. Zhu, B. T. Hu,
W. Zuo, J. Q. Li, Chin. Phys. Lett. {\bf 26}, 072301 (2009).
\bibitem{Zh10} H. F. Zhang, Y. J. Wang, J. M. Dong, J. Q. Li and W. Scheid, J. Phys. {\bf G 37}, 085107 (2010).
\bibitem{BV72} D. Vautherin and D. M. Brink, Phys. Rev. {\bf C 5}, 626 (1972).
\bibitem{BV75} Y. M. Engel, D. M. Brink, K. Goeke, S. J. Krieger and D. Vautherin, Nucl. Phys. {\bf A 249}, 215 (1975).
\bibitem{BV81} P. Bonche and D. Vautherin, Nucl. Phys. {\bf A 372}, 496 (1981).
\bibitem{bka2005} B. K. Agrawal, S. Shlomo, and V. Kim Au, {\it Phys. Rev. C} {\bf 72}, 014310 (2005).  
\bibitem{Ke35} E. C. Kemble, Phys. Rev. {\bf 48}, 549 (1935).
\bibitem{Ch97} E. Chabanat, P. Bonche, P. Haensel, J. Meyer and R. Schaeffer, {\it Nucl. Phys.} {\bf A 627}, 710 (1997).
\bibitem{Ch98} E. Chabanat, P. Bonche, P. Haensel, J. Meyer and R. Schaeffer, {\it Nucl. Phys.} {\bf A 635}, 231 (1998).
\bibitem{Ch981} E. Chabanat, P. Bonche, P. Haensel, J. Meyer and R. Schaeffer, {\it Nucl. Phys.} {\bf A 643}, 441 (1998).
\bibitem{Do72} C. B. Dover and N. V. Giai, Nucl. Phys. {\bf A 190}, 373 (1972).
\bibitem{dutra12}
M. Dutra , O. Lourenco, J. S. S. Martins , A. Delfino , J. R. Stone and P. D. Stevenson {\it Phys. Rev. } {\bf C 85}, 035201 (2012).
\bibitem{demorest2010}
 P. B. Demorest, T. Pennucci ,S. M. Ransom , M. S. E. Roberts \& J. W. T. Hessels, Nature, {\bf 467}, 1081(2010).
\bibitem{km17}
K. Madhuri, D.N. Basu, T.R. Routray and S.P. Pattnaik {\it Eur.Phys.J. } {\bf A 53}, 151 (2017).
\bibitem{stone2013}
P. D. Stevenson, P. M. Goddard, J. R. Stone and M. Dutra, {\it arXiv:1210.1592, AIP Conf. Proc.} {\bf 1529}, 262 (2013).
\bibitem{So02} A. A. Sonzogni, Nucl. Data Sheets {\bf 95}, 1 (2002).
\bibitem{Wo04} P. J. Woods et al., Phys. Rev. {\bf C 69}, 051302(R) (2004).
\bibitem{Jo06} D. T. Joss et al., Phys. Lett. {\bf B 641}, 34 (2006).

\end{thebibliography}
\end{document}